\documentclass[9pt,compsoc]{IEEEtran}

\usepackage{pgfplots}
\usepgfplotslibrary{groupplots}
\pgfplotsset{compat=1.17}
\usepackage{pgfplotstable}
\usepackage{graphicx}
\usepackage{subfig}
\usepackage{multirow}
\usepackage{lipsum}
\usepackage{xspace}
\usetikzlibrary{patterns}
\usepackage{amsmath}
\usepackage{amsfonts}
\usepackage{amssymb}

\newcommand*\circled[1]{\tikz[baseline=(char.base)]{
            \node[shape=circle,draw,inner sep=0.7pt] (char) {#1};}}
\usepackage{hyperref}
\hypersetup{colorlinks,allcolors=black}
\usepackage{float}
\usetikzlibrary{positioning}

\usepackage[usestackEOL]{stackengine}

\usepackage{pifont}
\ifCLASSOPTIONcompsoc
  \usepackage[nocompress]{cite}
\else
  \usepackage{cite}
\fi

\ifCLASSINFOpdf
\else
\fi

\begin{document}
\pagenumbering{gobble} 
\title{Scalability Bottlenecks in Multi-Agent Reinforcement Learning Systems}
%
%
%
%

\author{Kailash Gogineni, 
        Peng Wei, 
        Tian Lan and 
        Guru Venkataramani\\
The George Washington University, Washington, DC, USA\\
E-mail: \{kailashg26, pwei, tlan, guruv\}@gwu.edu}
\IEEEtitleabstractindextext{%
\begin{abstract}
Multi-Agent Reinforcement Learning (MARL) is a promising area of research that can model and control multiple, autonomous decision-making agents. During online training, MARL algorithms involve performance-intensive computations such as exploration and exploitation phases originating from large observation-action space belonging to multiple agents. In this article, we seek to characterize the scalability bottlenecks in several popular classes of MARL algorithms during their training phases. Our experimental results reveal new insights into the key modules of MARL algorithms that limit the scalability, and outline potential strategies that may help address these performance issues. 
\end{abstract}

}
\makeatletter
\def\endthebibliography{%
  \def\@noitemerr{\@latex@warning{Empty `thebibliography' environment}}%
  \endlist
}
\makeatother

\maketitle

\IEEEdisplaynontitleabstractindextext

%
\IEEEpeerreviewmaketitle

\section{Introduction}
\label{sec:introduction}
Reinforcement Learning~(RL) algorithms have widespread applications in robotics, aviation, autonomous driving, gaming, recommendation systems and healthcare. RL frameworks optimize AI agent behavior and its interactions with an environment by taking actions based on current observation/state space, evaluating the quality of state-action pairs using a reward function, and then transitioning to a new state~\cite{sutton2018reinforcement}. The function that determines the action is known as a policy. The agent aims to find an optimal policy that maximizes the total accumulative~(discounted) reward. The function representing the reward estimates is known as the value function. 
  
Multi-agent Reinforcement Learning~(MARL~\cite{sutton2018reinforcement}) is a rapidly growing research area where there is significant sharing of observations between the agents during training, and joint actions among these agents could affect the environment dynamically. Agents are trained to reach the goals while minimizing interference~(obstacles) with each other to perform competitive~\textcolor{black}{(e.g., Predator prey) and cooperative~(e.g., Cooperative navigation) tasks~\cite{lowe2017multi}}. 
In the cooperative setting, all the observations are shared and the training can be performed centrally. A competitive setting differentiates the agent pool- i.e., each agent aims to outperform its opponents. The actions taken by the individual agents affect other agents' behavior and their rewards dynamically in this environment. As a result, MARL training involves several {\it computationally-challenging} tasks that deal with dynamically changing environments. 
 \begin{figure}
    [!htb]\centering
    \includegraphics[scale=0.46]{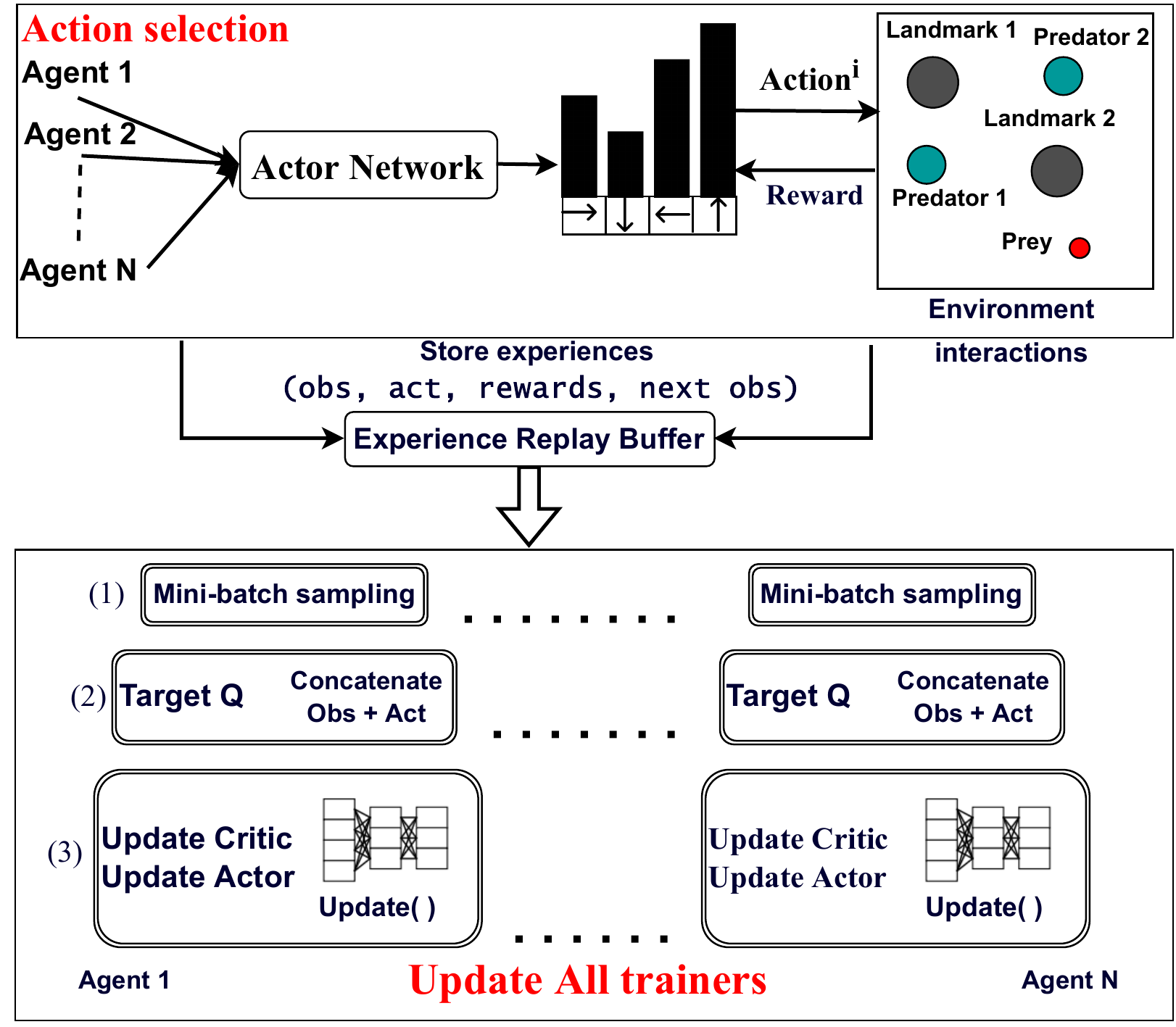}
    \caption{\textcolor{black}{Overview of our multi-agent decentralized actor, centralized critic approach~(Competitive environment).}}
  \label{fig-arch}
  \end{figure}

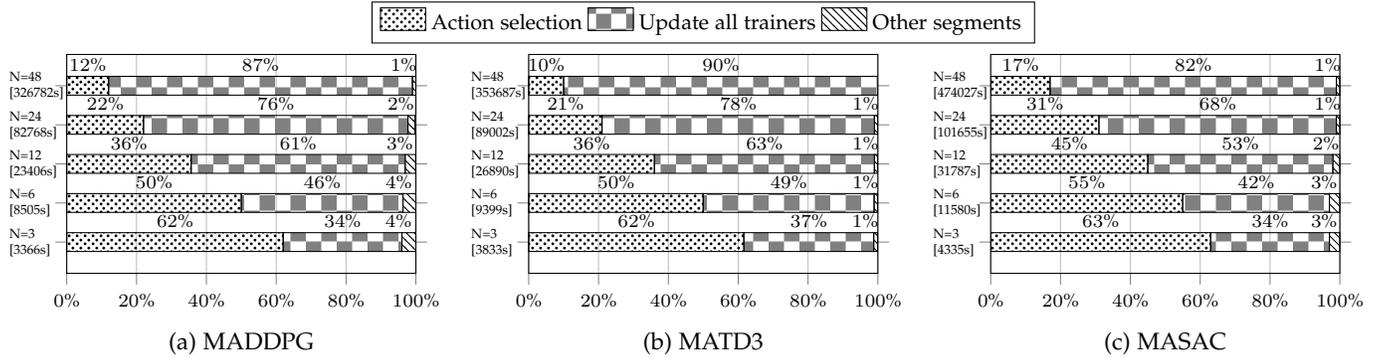
\begin{figure*}[ht]
\begin{tikzpicture}
\begin{groupplot}[
group style={group size=3 by 1, horizontal sep=1.5cm}, 
width=0.30\textwidth, height=0.3cm,
xmajorgrids=true,
tick align=outside, xtick pos=left,
scale only axis,
x post scale=0.85,
enlarge y limits=0.2,
xmin=0, xmax=100,
xtick={0,20,...,100},
ytick=data, 
xticklabel={\pgfmathprintnumber{\tick}\%},
xticklabel style={font=\scriptsize},
y tick label style={font=\tiny,anchor=east,align=left,text width=0.5cm,rotate=360},
nodes near coords style={font=\scriptsize,  yshift=8pt},
legend style={at={(0.5,1.05)}, anchor=south, legend columns=4, font=\footnotesize, cells={anchor=west}},
xbar stacked,
/pgf/bar width=2.5mm,
y=5.2mm,
legend image code/.code={%
            \draw[#1] (0cm,-0.1cm) rectangle (0.6cm,0.2cm);
            },
title style={at={(0.5,-0.3)}, anchor=north},
nodes near coords={\pgfkeys{/pgf/number format/precision=0}\pgfmathprintnumber{\pgfplotspointmeta}\%},
coordinate style/.condition={x-0.5*rawx>95}{xshift=-4pt},
]

\nextgroupplot[title=(a) MADDPG,
yticklabels={{N=3 \newline[3366s]},{N=6 \newline[8505s]},{N=12 \newline[23406s]},{N=24 \newline[82768s]},{N=48 \newline[326782s]}}]
\addplot [color=black, pattern color=black, pattern=crosshatch dots, %
nodes near coords={\pgfmathparse{less(\pgfplotspointmetatransformed,100)}%
\ifnum\pgfmathresult=1%
\hspace{8pt}%
\fi%
\pgfkeys{/pgf/number format/precision=0}\pgfmathprintnumber{\pgfplotspointmeta}\%}]
coordinates{(62,1) (50,2) (35.64,3) (22,4) (12,5)};
\addplot [color=black,  pattern color=gray, pattern=checkerboard] coordinates{(34,1) (46.3,2)(61.29,3) (75.72,4) (87,5)};
\addplot [color=black,  pattern color=black, pattern=north west lines] coordinates{(4,1) (3.7,2) (3,3) (2,4) (1,5)};
\nextgroupplot[title=(b) MATD3,
yticklabels={{N=3 \newline[3833s]},{N=6 \newline[9399s]},{N=12 \newline[26890s]},{N=24 \newline[89002s]},{N=48 \newline[353687s]}}]
\addplot [color=black,  pattern color=black, pattern=crosshatch dots, %
nodes near coords={\pgfmathparse{less(\pgfplotspointmetatransformed,100)}%
\ifnum\pgfmathresult=1%
\hspace{8pt}%
\fi%
\pgfkeys{/pgf/number format/precision=0}\pgfmathprintnumber{\pgfplotspointmeta}\%}]
coordinates{(61.62,1) (49.97,2) (36,3) (21,4) (10,5)};
\addplot [color=black,  pattern color=gray, pattern=checkerboard] 
coordinates{(37.20,1) (48.96,2) (63,3) (78,4) (90,5)};
\addplot [color=black, pattern color=black, pattern=north west lines] 
coordinates{(1.12,1) (1,2) (1,3) (1,4) (0,5)};
\legend{Action selection, Update all trainers, Other segments},

\nextgroupplot[title=(c) MASAC,
yticklabels={{N=3 \newline[4335s]},{N=6 \newline[11580s]},{N=12 \newline[31787s]},{N=24 \newline[101655s]},{N=48 \newline[474027s]}}]
\addplot [color=black,  pattern color=black, pattern=crosshatch dots]
coordinates{(63,1) (55,2) (45,3) (31,4) (17,5)};
\addplot [color=black,   pattern color=gray, pattern=checkerboard] coordinates{(34,1) (42,2)(53,3) (68,4) (82,5)};
\addplot [color=black, pattern color=black, pattern=north west lines] 
coordinates{(3,1) (3,2) (2,3) (1,4) (1,5)};

\end{groupplot}
\end{tikzpicture}
\caption{Training time breakdown for three MARL workloads with 3 to 48 agents. The environment is Competitive task~(Predator-Prey). The total training time of MARL algorithms~(in seconds) is shown on y-axis within square brackets.}
\label{figure1}
\end{figure*}

In this article, we seek to understand the {\it key scalability bottlenecks} on well-known model-free MARL frameworks~\cite{lowe2017multi, ackermann2019reducing, haarnoja2018soft} implemented using actor-critic methods 
\textcolor{black}{with state spaces that are usually very large}. 
We analyze different MARL training phases where the actor and critic networks are responsible for policy and value functions, respectively. 
As shown in Figure~\ref{fig-arch}, the actor network outputs the actions when a group of agents interact with the environment and each agent learns an individual policy that maps its observations to optimal actions~(\textit{Action selection}) to maximize the expected return. 
During the \textit{mini-batch sampling} phase, each agent $i$ collects the observations, actions, and new observations of all other agents stored within the \textit{Experience Replay Buffer}. The critic updates the value function using random mini-batch samples of all the agents obtained from the past experiences stored in the replay buffer. 
During \textit{Update all trainers} phase, the actor network is updated using Q-values computed by the critic~\cite{lowe2017multi}.
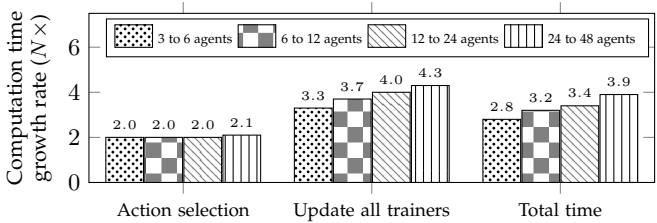
\begin{figure}[h]
    \centering
    \pgfplotstableread[col sep=comma]{
X,  Y1, Y2, Y3, Y4
Action selection, 2.0, 2, 2, 2.1
Update all trainers,  3.3, 3.7, 4.0, 4.3
Total time, 2.8, 3.2, 3.4, 3.9
}\mydata
\begin{tikzpicture}
\begin{axis}[
height=50mm, width=0.5\textwidth,
bar width=0.20,
ybar=0.5pt,
enlarge y limits=0,
enlarge x limits={abs=0.5},
ymin=0,
ymax=7.5,
y=3.0mm,
legend style={at={(0.5,1.05)}, anchor=south, legend columns=4, font=\footnotesize, cells={anchor=west}},
legend image code/.code={%
            \draw[#1] (0cm,-0.1cm) rectangle (0.5cm,0.2cm);
            },
x tick label style = {font = \scriptsize, text width = 2.4cm, align = center, rotate = 360, anchor = north},
ylabel style={align=center, font=\footnotesize}, ylabel={Computation time\\growth rate ($N\times$)},
xtick=data,
xticklabels from table={\mydata}{X},
nodes near coords,
nodes near coords style={font=\tiny,
/pgf/number format/.cd,
precision=1,
zerofill,
},
legend style={legend pos=north west,
cells={anchor=west},
font=\tiny,
}
]
\addplot [ybar, pattern color=black, pattern=crosshatch dots] table [x expr=\coordindex,y=Y1]{\mydata};
\addplot [ybar, pattern color=gray, pattern=checkerboard] table [x expr=\coordindex,y=Y2]{\mydata};
\addplot [ybar, pattern color=gray, pattern=north west lines] table [x expr=\coordindex,y=Y3]{\mydata};
\addplot [ybar, pattern color=black, pattern=vertical lines] table [x expr=\coordindex,y=Y4]{\mydata};
\legend{3 to 6 agents, 6 to 12 agents, 12 to 24 agents, 24 to 48 agents}
\end{axis}
\end{tikzpicture}
    \caption{Computation time growth in MARL modules averaged across three MARL frameworks.}
    \label{Figure4}
\end{figure}

\section{Motivation}
\label{sec:motivation}

MARL training is performance-intensive as the agents' policies continually evolve, and the replay buffer samples will be refreshed to find an optimal policy for the inference~\cite{lowe2017multi}. 
Figure~\ref{figure1} shows that \textit{Update all trainers} contributes to $\approx$35\% to $\approx$90\% of the training time as the number of MARL agents grow from 3 to 48. This is primarily due to two reasons:~\circled{1}~In MARL, each agent has its own actor and critic networks since they may have different rewards. Each agent has to randomly sample a mini-batch of transitions from the replay buffer to update both the critic and actor networks. This requires each agent to sample experiences from every other agent. 
\circled{2} The dynamic memory requirements of observation and action spaces also grow quadratically due to each agent having to coordinate with other agents towards sharing their observations and actions. 
We observe that other MARL phases, such as \textit{Action selection} occupy a small portion and scales linearly with the number of agents~(Figure~\ref{Figure4}). 
This is because, action selection is performed with individual agents' policy using local observations and interactions with the environment. \textit{Other segments} is a combination of experience collection, reward collection and policy initialization and they add a negligible overhead.~\textcolor{black}{Note that the agents interact in a shared environment, and the training time is summed over all agents}.

\textcolor{black}{Prior studies, like FA3C~\cite{cho2019fa3c}, have focused on accelerating multiple parallel worker scenarios, where each agent is controlled independently within their own environments using single-agent RL algorithms. 
In contrast, we seek to understand multi-agent learning frameworks, where the agents operate in a single shared environment. 
Agents in such MARL settings usually have high visibility of one another (leading to large space and action spaces). 
To the best of our knowledge, this is the first characterization study of MARL scalability bottlenecks. 
}
\section{Background}
\label{sec:Background and Related work}

Typically, MARL settings with $N$ agents is defined by a set of states, $S = S_{1} \times ... \times S_{N}$, a set of actions $A = A_{1} \times ... \times A_{N}$. 
 Each agent selects its action by using a policy $\pi_{\theta_{i}} : O_{i} \times A_{i} \rightarrow [0, 1]$. The state transition~($T : S \times A_{1} \times A_{2} \times ... \times A_{N}$) function produces the next state $S^{'}$, given the current state and actions for each agent. The reward, $R_{i} : S \times A_{i} \rightarrow \mathbb{R}$ for each agent is a function of global state and action of \textit{all other agents}, with the aim of maximizing its own expected return $R_{i} = \sum_{t=0}^{T} \gamma^{t}r_{i}^{t}$, where $\gamma$ denotes the discount factor and $T$ is the time horizon. For this, we use the actor-critic methods such as MADDPG~\cite{lowe2017multi}, MATD3~\cite{ackermann2019reducing}, MASAC~\cite{haarnoja2018soft}. 
 
 In \textit{MADDPG}~\cite{lowe2017multi}, each agent learns an individual policy that maps the observation to its action to maximize the expected return, which is approximated by the critic. MADDPG lets the critic of agent $i$ to be trained  by minimizing the loss with the \textit{target Q-value} and $y_i$ using $\mathcal{L}(\theta_{i}) = {\rm I\!E}_{D}[(Q_{i}(S,A_{1},...A_{n}) - y_{i}^{2}]$, and  $y_{i}=r_{i} + \gamma \overline Q_{i}(S^{'},A_{1}^{'},...A_{n}^{'})_{a_{j}^{'}=\overline \pi(o_{j}^{'})}$, where $S$ and $A_{1},...A_{n}$ represent the joint observations and actions respectively. $D$ is the experience replay buffer that stores the \textit{observations, actions, rewards, and new observations} samples of all agents obtained after the training episodes. The MARL framework has four networks- actor, critic, target actor, and target critic. $\overline Q_{i}$ and $\overline \pi(o_{j}^{'})$ are the target networks for the stable learning of critic~($Q_{i}$) and actor networks. The target actor estimates next action from the policy using the state output by the actor network. The target critic aggregates the output from the target actor to compute the target Q-values, that helps to update the critic network and assess the quality of of actions taken by agents. The target networks are created to achieve training stability. Note that the updating sequence of networks in the back-propagation phase is critics, actors, then the target networks.
 
 \textit{MATD3}~\cite{ackermann2019reducing} uses the twin delayed critics to tackle the over-estimation bias problem~\cite{ackermann2019reducing} and incorporates small amounts of noise to the actions sampled from the buffer. As the change of critic values needs to be reflected in the policies of other agents, MATD3 employs delayed policy updates for target networks and the policies to obtain an accurate critic before using it to update the actor. In the domains where it is necessary to learn a winning strategy~(e.g., Predator-Prey, Cooperative navigation), MATD3 outperforms MADDPG~\cite{ackermann2019reducing}. 
 
\textit{MASAC}~\cite{haarnoja2018soft} improves the convergence properties over MADDPG and MATD3, and obtains higher returns in the competitive and cooperative environments. MASAC uses the maximum entropy RL, in which the agents are encouraged to maximize the exploration within the policy. MASAC assigns equal probability to nearly-optimal actions which have similar state-action values and avoids repeatedly selecting the same action. This learning trick will increase the stability, policy exploration and the sample efficiency~\cite{iqbal2019actor, haarnoja2018soft}.

\begin{table}[h]
\caption{\label{table-1}Multi-agent Particle environment.}
\begin{tabular}{
  |p{\dimexpr.23\linewidth-2\tabcolsep-1.3333\arrayrulewidth}
  |p{\dimexpr.77\linewidth-2\tabcolsep-1.3333\arrayrulewidth}|
  }
\hline
\textbf{Environment} & \textbf{Details}\\
\hline
Cooperative navigation & \textit{N} agents move in a cooperated manner to reach \textit{L} landmarks and the rewards encourages the agents get closer to the landmarks.\\
\hline
Predator-Prey & \textit{N} predators work cooperatively to block the way of \textit{M} fast paced prey agents. The prey agents are environment controlled and they try to avoid the collision with predators.\\
\hline
\end{tabular}
\end{table}
\section{Evaluation Setup}\label{sec:Evaluation_setup}
We evaluate three state-of-the-art MARL algorithms, MADDPG, MATD3, and MASAC using Multi-agent Particle Environment~(MPE~\cite{lowe2017multi}). 
We outline the behavior of the selected environments in Table~\ref{table-1}. 
The actor and critic networks are paramterized by a two-layer ReLU MLP with 64 units per layer
and mini-batch size is 1024 for sampling the transitions. In all of our experiments, we use Adam optimizer~\cite{kingma2014adam} with a learning rate of 0.01, maximum episode length as 25 (max episodes to reach the terminal state) and $\tau$ = 0.01 for updating the target networks. $\gamma$ is the discount factor which is set to 0.95. The size of replay buffer is $10^{6}$ and entropy coefficient for MASAC is 0.05. The network parameters are updated after every 100 samples added to the replay buffer.

\textcolor{black}{All the workloads are trained and profiled on Nvidia GeForce~RTX~3090 Ampere Architecture connected with AMD Ryzen Threadripper PRO 3975WX CPU, which has 32 cores with 128 MiB of Last-Level Cache, 512 Gigabytes of main memory and the CPU's clock speed of 3.5GHz. The server runs on Ubuntu Linux~20.04.5~LTS operating system with CUDA~9.0, cuDNN~7.6.5, PCIe Express®~v4.0 with 
NCCL~v2.8.4 communication library. The machine supports python 3.7.15,~Tensorflow (v2.11.0), Tensorflow-GPU~(v2.1.0) and OpenAI GYM~(v0.10.5). We use Perf~\cite{perf} tool and
hardware performance counters for performance analysis. The workloads are trained for 60K episodes using default hyper-parameters recommended by the algorithms.}

\section{Experimental Evaluation}
\label{sec:Observation and analysis}

For deeper analysis, we divide \textit{Update all trainers} into multiple modules: \textit{Mini-batch sampling, Target Q calculation}, and \textit{Q loss \& P loss} and present our results in the primarily competitive setting (predator-prey) in order to understand the key factors limiting MARL scalability. We note that the predator agents operate in the cooperative setting to maximize their shared return. Therefore, our test-bed allows us to evaluate both competitive and cooperative agent scenarios.

\begin{figure*}[!htpb]
\begin{tikzpicture}
\begin{groupplot}[
group style={group size=3 by 1, horizontal sep=1.5cm}, 
width=0.30\textwidth, height=0.3cm,
xmajorgrids=true,
tick align=outside, xtick pos=left,
scale only axis,
x post scale=0.85,
enlarge y limits=0.20,
xmin=0, xmax=100,
xtick={0,20,...,100},
ytick=data, 
xticklabel={\pgfmathprintnumber{\tick}\%},
xticklabel style={font=\scriptsize},
y tick label style={font=\tiny,anchor=east,align=left,text width=0.5cm,rotate=360},
nodes near coords style={font=\tiny,  yshift=8pt},
legend style={at={(0.5,1.05)}, anchor=south, legend columns=4, font=\footnotesize, cells={anchor=west}},
xbar stacked,
/pgf/bar width=2.5mm,
y=5.2mm,
legend image code/.code={%
            \draw[#1] (0cm,-0.1cm) rectangle (0.6cm,0.2cm);
            },
title style={at={(0.5,-0.3)}, anchor=north},
nodes near coords={\pgfkeys{/pgf/number format/precision=0}\pgfmathprintnumber{\pgfplotspointmeta}\%},
coordinate style/.condition={x-0.5*rawx>95}{xshift=-4pt},
]
\nextgroupplot[title=(a) MADDPG,
yticklabels={{N=3 \newline[1144s]},{N=6 \newline[3912s]},{N=12 \newline[14278s]},{N=24 \newline[62904s]},{N=48 \newline[284299s]}}]
\addplot [color=black, pattern color=black, pattern=crosshatch dots, %
nodes near coords={\pgfmathparse{less(\pgfplotspointmetatransformed,100)}%
\ifnum\pgfmathresult=1%
\hspace{16pt}%
\fi%
\pgfkeys{/pgf/number format/precision=0}\pgfmathprintnumber{\pgfplotspointmeta}\%}] coordinates{(59.08,1) (64,2) (65,3) (65,4) (64,5)};
\addplot [color=black,  pattern color=gray, pattern=checkerboard, %
nodes near coords={\pgfmathparse{less(\pgfplotspointmetatransformed,100)}%
\ifnum\pgfmathresult=1%
\hspace{-8pt}%
\fi%
\pgfkeys{/pgf/number format/precision=0}\pgfmathprintnumber{\pgfplotspointmeta}\%}] coordinates{(17.69,1) (19,2) (21,3) (23,4) (24,5)};
\addplot [color=black,pattern color=gray, pattern=north east lines, %
nodes near coords={\pgfmathparse{less(\pgfplotspointmetatransformed,100)}%
\ifnum\pgfmathresult=1%
\hspace{-10pt}%
\fi%
\pgfkeys{/pgf/number format/precision=0}\pgfmathprintnumber{\pgfplotspointmeta}\%}] coordinates{(10.69,1) (9,2)(8,3) (6,4) (6,5)};
\addplot [color=black,  pattern color=black, pattern=north west lines, %
nodes near coords={\pgfmathparse{less(\pgfplotspointmetatransformed,100)}%
\ifnum\pgfmathresult=1%
\hspace{8pt}%
\fi%
\pgfkeys{/pgf/number format/precision=0}\pgfmathprintnumber{\pgfplotspointmeta}\%}] coordinates{(12.08,1) (8,2) (6,3) (6,4) (6,5)};

\nextgroupplot[title=(b) MATD3,
yticklabels={{N=3 \newline[1418s]},{N=6 \newline[4606s]},{N=12 \newline[16941s]},{N=24 \newline[69422s]},{N=48 \newline[318319s]}}]
\addplot [color=black,pattern color=black, pattern=crosshatch dots
, %
nodes near coords={\pgfmathparse{less(\pgfplotspointmetatransformed,100)}%
\ifnum\pgfmathresult=1%
\hspace{16pt}%
\fi%
\pgfkeys{/pgf/number format/precision=0}\pgfmathprintnumber{\pgfplotspointmeta}\%}]
coordinates{(56,1) (60,2) (61,3) (61,4) (61,5)};
\addplot [color=black,  pattern color=gray, pattern=checkerboard
,
nodes near coords={\pgfmathparse{less(\pgfplotspointmetatransformed,100)}%
\ifnum\pgfmathresult=1%
\hspace{-8pt}%
\fi%
\pgfkeys{/pgf/number format/precision=0}\pgfmathprintnumber{\pgfplotspointmeta}\%}]  
coordinates{(18,1) (20,2) (22,3) (24,4) (25,5)};
\addplot [color=black, pattern color=gray, pattern=north east lines
, %
nodes near coords={\pgfmathparse{less(\pgfplotspointmetatransformed,100)}%
\ifnum\pgfmathresult=1%
\hspace{-10pt}%
\fi%
\pgfkeys{/pgf/number format/precision=0}\pgfmathprintnumber{\pgfplotspointmeta}\%}] 
coordinates{(15,1) (12,2)(10,3) (9,4) (9,5)};
\addplot [color=black,  pattern color=black, pattern=north west lines
, %
nodes near coords={\pgfmathparse{less(\pgfplotspointmetatransformed,100)}%
\ifnum\pgfmathresult=1%
\hspace{8pt}%
\fi%
\pgfkeys{/pgf/number format/precision=0}\pgfmathprintnumber{\pgfplotspointmeta}\%}] 
coordinates{(11,1) (8,2) (7,3) (6,4) (5,5)};
\legend{Mini-batch sampling, Target Q calculation, Q loss, P loss},

\nextgroupplot[title=(c) MASAC,
yticklabels={{N=3 \newline[1474s]},{N=6 \newline[4865s]},{N=12 \newline[16848s]},{N=24 \newline[69127s]},{N=48 \newline[388700s]}}]
\addplot [color=black, pattern color=black, pattern=crosshatch dots
, %
nodes near coords={\pgfmathparse{less(\pgfplotspointmetatransformed,100)}%
\ifnum\pgfmathresult=1%
\hspace{16pt}%
\fi%
\pgfkeys{/pgf/number format/precision=0}\pgfmathprintnumber{\pgfplotspointmeta}\%}] 
coordinates{(58,1) (62,2) (63,3) (63,4) (62,5)};
\addplot [color=black,  pattern color=gray, pattern=checkerboard
, %
nodes near coords={\pgfmathparse{less(\pgfplotspointmetatransformed,100)}%
\ifnum\pgfmathresult=1%
\hspace{-8pt}%
\fi%
\pgfkeys{/pgf/number format/precision=0}\pgfmathprintnumber{\pgfplotspointmeta}\%}] 
coordinates{(19,1) (21,2) (23,3) (24,4) (25,5)};
\addplot [color=black, pattern color=gray, pattern=north east lines 
, %
nodes near coords={\pgfmathparse{less(\pgfplotspointmetatransformed,100)}%
\ifnum\pgfmathresult=1%
\hspace{-10pt}%
\fi%
\pgfkeys{/pgf/number format/precision=0}\pgfmathprintnumber{\pgfplotspointmeta}\%}] 
coordinates{(12,1) (10,2)(8,3) (7,4) (7,5)};
\addplot [color=black,  pattern color=black, pattern=north west lines 
, %
nodes near coords={\pgfmathparse{less(\pgfplotspointmetatransformed,100)}%
\ifnum\pgfmathresult=1%
\hspace{3pt}%
\fi%
\pgfkeys{/pgf/number format/precision=0}\pgfmathprintnumber{\pgfplotspointmeta}\%}] 
coordinates{(11,1) (7,2) (6,3) (6,4) (6,5)};


\end{groupplot}
\end{tikzpicture}
\caption{Training time breakdown within \textit{Update all trainers} on three different MARL workloads with 3 to 48 agents under Predator-Prey environment.~The total training time of \textit{Update all trainers}~(in seconds) is shown on y-axis within square brackets.}
\label{figure2}
\end{figure*}
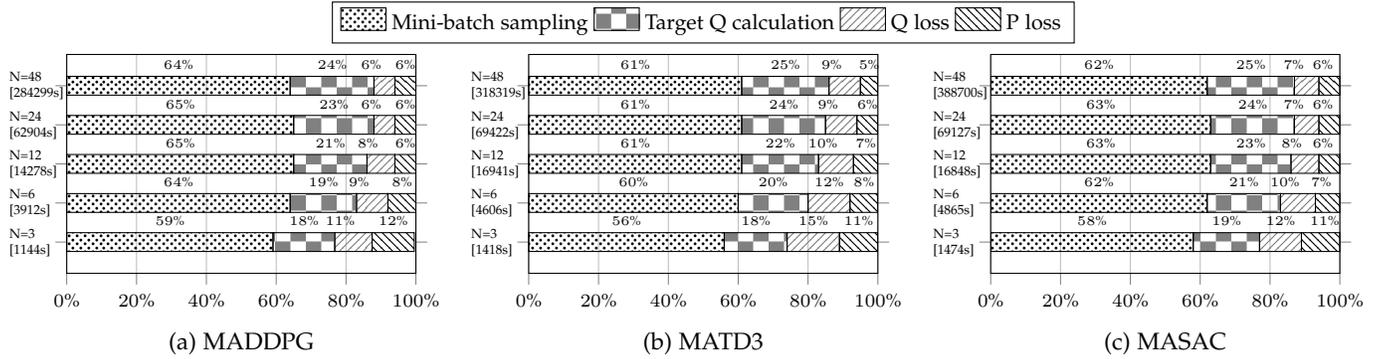

 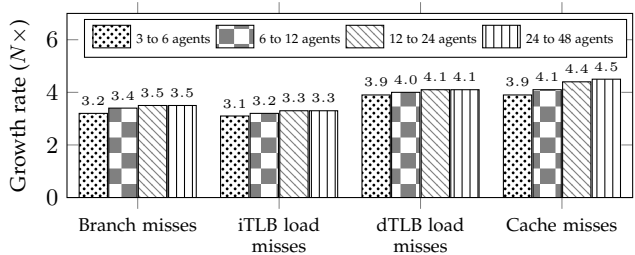
\begin{figure}[b]
    \centering
    \pgfplotstableread[col sep=comma]{
X,  Y1, Y2, Y3, Y4
Branch misses, 3.2, 3.4, 3.5, 3.5 
iTLB load misses, 3.1, 3.2, 3.3, 3.3
dTLB load misses,  3.9, 4.0, 4.1,4.1 
Cache misses, 3.9, 4.1, 4.4, 4.5
}\mydata
\begin{tikzpicture}
\begin{axis}[
height=50mm, width=0.5\textwidth,
bar width=0.20,
ybar=0.5pt,
enlarge y limits=0,
enlarge x limits={abs=0.5},
ymin=0,
ymax=7,
y=3.5mm,
legend style={at={(0.5,1.05)}, anchor=south, legend columns=4, font=\footnotesize, cells={anchor=west}},
legend image code/.code={%
            \draw[#1] (0cm,-0.1cm) rectangle (0.5cm,0.2cm);
            },
x tick label style = {font = \scriptsize, text width = 1.8cm, align = center, rotate = 360, anchor = north},
ylabel style={align=center, font=\footnotesize}, ylabel={Growth rate ($N\times$)},
xtick=data,
xticklabels from table={\mydata}{X},
nodes near coords,
nodes near coords style={font=\tiny,
/pgf/number format/.cd,
precision=1,
zerofill,
},
legend style={legend pos=north west,
cells={anchor=west},
font=\tiny,
}
]
\addplot [ybar, pattern color=black, pattern=crosshatch dots] table [x expr=\coordindex,y=Y1]{\mydata};
\addplot [ybar, pattern color=gray, pattern=checkerboard] table [x expr=\coordindex,y=Y2]{\mydata};
\addplot [ybar, pattern color=gray, pattern=north west lines] table [x expr=\coordindex,y=Y3]{\mydata};
\addplot [ybar, pattern color=black, pattern=vertical lines] table [x expr=\coordindex,y=Y4]{\mydata};
\legend{3 to 6 agents, 6 to 12 agents, 12 to 24 agents, 24 to 48 agents}
\end{axis}
\end{tikzpicture}
\caption{\textcolor{black}{Hardware performance analysis of \textit{Update all trainers} averaged across three MARL workloads when the number of agents are scaled by 2$\times$.~Performance analysis is averaged across three sub-functions of \textit{Update all trainers}.}}
    \label{Figurenew}
\end{figure}

\textbf{Overview of Profile.}\textcolor{black}{~Figure~\ref{figure2} shows the breakdown between the modules, \textit{Mini-batch sampling, Target Q calculation, Q loss and P loss} that contribute 61\%, 21\%, 10\%, and 8\% to computation time averaging across different workloads respectively. With increasing number of agents from 3 to 48, we observe that Instructions Per Cycle (IPC) 
steadily drops by over 21\%~(1.37 to 1.08), and global cache misses increase by 54\%, which indicates that the instruction throughput and cache performance slumps when more agents are involved.~A super-linear growth rate of certain key architectural features like branch misses, TLB load misses, and global cache misses further validate the presence of computational bottlenecks across MARL workloads (Figure~\ref{Figurenew}).}
\begin{figure}[ht]
    \raggedright
\begin{tikzpicture}
    \begin{groupplot}[
group style={
    group size=3 by 1,
    ylabels at=edge left,
    horizontal sep=5mm,
            },
    width=0.44\linewidth,
    enlarge x limits=0.3,
    title style = {yshift=-1ex, font=\scriptsize, align=center},
    xlabel={Agents~$(N \to \ 2N)$},
    ylabel={Growth rate~($N\times$)},
    label style = {font=\scriptsize},
    symbolic x coords={3-6, 6-12, 12-24, 24-48},
    xtick={3-6, 6-12, 12-24, 24-48},
    ticklabel style={font=\tiny},
    ymin=0,    ymax=7,
    nodes near coords,
    every node near coord/.append style={font=\scriptsize},
    nodes near coords align={vertical},
                    ]
\nextgroupplot[title={(a) Mini-batch sampling}]
        \addplot[ybar, pattern color=black, pattern=crosshatch dots] coordinates {  (3-6, 3.5)};
        \addplot[ybar, pattern color=gray, pattern=checkerboard] coordinates { (6-12, 3.7)};
        \addplot[ybar, pattern color=gray, pattern=north east lines] coordinates {  (12-24,4.0)};
        \addplot[ybar, pattern color=gray, pattern=horizontal lines] coordinates {  (24-48,4.2)};
\nextgroupplot[title={(b) Target Q Calculation}]
     \addplot[ybar, pattern color=black, pattern=crosshatch dots] coordinates {  (3-6, 3.8)};
        \addplot[ybar, pattern color=gray, pattern=checkerboard] coordinates { (6-12, 4.2)};
        \addplot[ybar,pattern color=gray, pattern=north east lines] coordinates {  (12-24, 4.5)};
        \addplot[ybar, pattern color=gray, pattern=horizontal lines] coordinates {  (24-48,4.7)};
\nextgroupplot[title={(c) Q loss \& P loss}]
`       \addplot[ybar, pattern color=black, pattern=crosshatch dots] coordinates {  (3-6, 2.6)};
        \addplot[ybar, pattern color=gray, pattern=checkerboard] coordinates { (6-12,3.2)};
        \addplot[ybar, pattern color=gray, pattern=north east lines] coordinates {  (12-24, 3.4)};
        \addplot[ybar, pattern color=gray, pattern=horizontal lines] coordinates {  (24-48,3.7)};
\end{groupplot}
        \end{tikzpicture}
\caption{Computation time growth rate of the sub-functions within \textit{Update all trainers} averaged across three MARL workloads when the number of agents are scaled by 2$\times$.} 
\label{Figure3}
\end{figure}
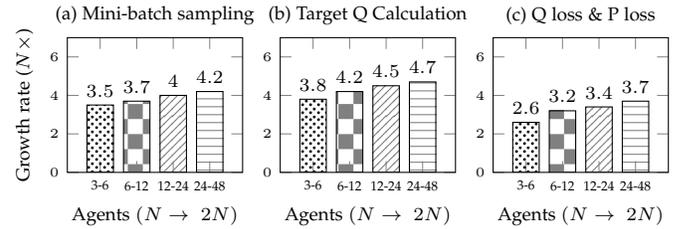
\subsection{Mini-batch sampling} 
Our experimental results in Figure~\ref{Figure3} shows super-linear increase in computation time with the number of agents during mini-batch sampling, the largest phase within the {\it Update All Trainers} module.
The is also consistently reflected in other related performance metrics: \textit{Total instructions}-$4\times$ and \textit{LLC-Load-misses}-$3\times$. The competitive behavior between prey agents and predator agents involves each predator agent receiving the velocity, position relative to all other predator agents and landmarks as observations to hunt the prey. A cooperative behavior also exists between the predator agents to maximize their shared return. Note that the agent replay buffers are kept separate from each other to capture their  individual past transitions. 

In this module, each agent has to randomly sample a set of mini-batch samples uniformly from others' replay buffers and update the parameters of its critic network. 
\textcolor{black}{Each agent $i$ performs \textit{lookup-read-write} operations, which grow as a function of number of agents, $N$ and this is repeated on all N agents. The time complexity to collect the transition set is $O(N^2K)$, $K$ being the batch size. 
} 
In cooperative navigation~(\textit{simple spread}~\cite{lowe2017multi}), we observe similar scalability bottlenecks since all the agents are trained together to reach the landmarks while avoiding collisions with each other.


\subsection{Target Q calculation}
The \textit{Target Q calculation} phase is second largest time consuming phase within {\it Update All Trainers}. Figure~\ref{Figure3} shows that this phase grows by $4\times$ with the number of agents. \textcolor{black}{Note that, in Figure~\ref{figure2}, the computation time as a percentage within {\it Update All Trainers} increases with the number of agents for \textit{target Q}, whereas the execution time proportion of \textit{Q loss and P loss} decrease slightly. This is because, the \textit{target Q} grows by a higher rate compared to \textit{Q loss}~(Figure~\ref{Figure3})}. 
Each agent performs the \textit{next action calculation, target Q next, and target Q values} as a function of all other agents' joint observation-action space. To calculate the \textit{next action}, the agent $i$ uses its policy network to determine \textit{next action-a'} from the \textit{next state-S'}. \textcolor{black}{In this phase, each agent's policy network involves multiplications with input-weight matrix and additions resulting in performance impact. The obtained a' and S' data are aggregated and concatenated into a single vector in order to compute the \textit{target Q next} amongst the cooperating agents. The input space~(dimension) for the \textit{Q-function} increases quadratically with the number of agents~\cite{sheikh2020multi}.}
The target critic values for each agent $i$ is computed using \textit{target Q next} values from the target actor network. We note that, each agent has to read other agents' policy values; as such for $N$ agents, there is $N\times(N-1)$ memory lookup operations corresponding to the \textit{next action-a'}. 
\subsection{Back-propagation - Q loss \& P loss} 


Back propagation is the third largest phase of {\it Update all Trainers}. 
This phase is dominated by the back-propagation of \textit{critic network} that computes the Mean-Squared Error loss between the target critic and critic networks, and the \textit{actor network} is updated by minimizing the Q values~(critic network). As the number of agents increases, the trainable parameters increase, and \textit{N} policy and \textit{N} critic networks are built for all N agents, which incurs extra time to update the weights for each agent. 
The average computation overhead of critic network for each iteration grows by up to 27\% for every $2\times$ increase in the agents. For the actor network, the average computation overhead grows by 36\%, which can help explain the performance bottlenecks involved in updating the weights for the individual agent networks. 



\section{Architectural Guidelines}
\label{sec:Architectural guidelines}

Architectural primitives implementing selective attention~\cite{iqbal2019actor} may help relieve some of the MARL scalability bottlenecks in \textit{target Q calculation and the critic network} toward reducing the input space for the networks. 

\textcolor{black}{
To address the data movement-related performance issues, we note that processing in- or near-memory accelerators may help improve the related architectural bottlenecks significantly. For example, recent GDDR6-based Processing-in-Memory~(PIM)~\cite{lee20221ynm} may be used to accelerate the matrix multiply-accummulate and activation operations for all of the four neural networks in most current MARL frameworks. 
Future PIM designs may also be augmented to incorporate MARL-specific hardware primitives (e.g., efficient mini-batch sampling) for improved hardware efficiency. Multi-core PIM Neural network accelerators can be used to exploit the higher coarse-grain hardware parallelism offered by multiple neural networks in MARL algorithms.} 

\textcolor{black}{
Code optimizations and data parallelism can also be leveraged during \textit{mini-batch sampling}. Shared memory multi-threading may be used to perform faster sampling (e.g., via OpenMP). This can help reduce the computational bottlenecks with each agent sampling its own transitions with parallel threads~\cite{kaler2022accelerating}. Our hardware performance analysis also shows the potential for improving the branch and TLB behavior, and the need for optimizations in hardware and software that can alleviate these issues.
}

For the input to critic networks, multi-level compression~\cite{jain2018gist}
techniques on selected group of agents may be used based on their importance in the environment. Also, LLC-Load-misses during mini-batch sampling are indicative of competition for the LLC cache, that may be addressed through smart cache allocation strategies. Other modules such as \textit{next action calculation, environment interactions, and action selection} phases may also benefit from  parallelization and custom acceleration of key modules. 

\section{Conclusion}
\label{sec:conclusion}
In this work, we present a detailed, end-to-end characterization of several popular Multi-Agent Reinforcement Learning algorithms and in particular,~explore the scalability bottlenecks in these workloads. Our experimental analysis present key insights on the modules that are driving factors behind scalability bottlenecks, and outline architectural guidelines to overcome them.  
\ifCLASSOPTIONcompsoc
  \section*{Acknowledgments}
\else
  \section*{Acknowledgment}
\fi
This research is based on work supported by the National Science Foundation under grant CCF-2114415.




%


\bibliographystyle{IEEEtran}
\bibliography{referencesold}

\end{document}